# Environmental versatility promotes modularity in genome-scale metabolic networks

**Areejit Samal[1,2], Andreas Wagner[3,4,5,#,\*], Olivier C. Martin[1,6,#,\*]**

[1] Laboratoire de Physique Théorique et Modèles Statistiques, CNRS and Univ Paris-Sud, UMR 8626, F-91405 Orsay Cedex, France
[2] Max Planck Institute for Mathematics in the Sciences, Inselstr. 22, 04103 Leipzig, Germany
[3] Institute of Evolutionary Biology and Environmental Studies, University of Zurich, Winterthurerstrasse 190, CH-8057 Zurich, Switzerland
[4] Swiss Institute of Bioinformatics, Quartier Sorge, Batiment Genopode, 1015 Lausanne, Switzerland
[5] The Santa Fe Institute, 1399 Hyde Park Road, Santa Fe, NM 87501, USA
[6] INRA, UMR 0320/UMR 8120 Génétique Végétale, Univ Paris-Sud, F-91190 Gif-sur-Yvette, France

\# Equal Contribution
\* Correspondence should be addressed to: Andreas Wagner or Olivier C. Martin

Email addresses:
    AS: samal@mis.mpg.de
    AW: andreas.wagner@ieu.uzh.ch
    OCM: olivier.martin@u-psud.fr

## Abstract

**Background:** The ubiquity of modules in biological networks may result from an evolutionary benefit of a modular organization. For instance, modularity may increase the rate of adaptive evolution, because modules can be easily combined into new arrangements that may benefit their carrier. Conversely, modularity may emerge as a by-product of some trait. We here ask whether this last scenario may play a role in genome-scale metabolic networks that need to sustain life in one or more chemical environments. For such networks, we define a network module as a maximal set of reactions that are fully coupled, *i.e.,* whose fluxes can only vary in fixed proportions. This definition overcomes limitations of purely graph based analyses of metabolism by exploiting the functional links between reactions. We call a metabolic network viable in a given chemical environment if it can synthesize all of an organism's biomass compounds from nutrients in this environment. An organism's metabolism is highly versatile if it can sustain life in many different chemical environments. We here ask whether versatility affects the modularity of metabolic networks.
**Results:** Using recently developed techniques to randomly sample large numbers of viable metabolic networks from a vast space of metabolic networks, we use flux balance analysis to study *in silico* metabolic networks that differ in their versatility. We find that highly versatile networks are also highly modular. They contain more modules and more reactions that are organized into modules. Most or all reactions in a module are associated with the same biochemical pathways. Modules that arise in highly versatile networks generally involve reactions that process nutrients or closely related chemicals. We also observe that the metabolism of *E. coli* is significantly more modular than even our most versatile networks.

**Conclusions:** Our work shows that modularity in metabolic networks can be a by-product of functional constraints, e.g., the need to sustain life in multiple environments. This organizational principle is insensitive to the environments we consider and to the number of reactions in a metabolic network. Because we observe this principle not just in one or few biological networks, but in large random samples of networks, we propose that it may be a generic principle of metabolic network organization.

## Introduction

The architectures of most multi-cellular organisms are strikingly modular. On the one hand, such modularity can be spatial. Organisms are partitioned into organs and tissues whose cells have specialized functions [1-2], and where cells of similar types are in close proximity. Such spatial modularity also exists within cells. Examples include organelles, spatial modules that allow specialized tasks to be performed in localized regions of a cell. Spatial modularity can be thought of as functional specialization according to spatial localization.

On the other hand, modularity can be topological, as research of the last ten years has shown. Such modularity is evident in biological networks such as protein-protein interaction networks [3-4], transcriptional regulatory networks [5], or metabolic networks [6-10]. In these systems, the networks – viewed as graphs – contain modules that are subsets of nodes strongly connected to each other but weakly connected to the remaining network. This kind of modularity does not involve explicit spatial location but nevertheless relies on a notion of proximity (of nodes in a network). If nodes within a module tend to be involved in the same biological or biochemical function, then both spatial and topological modularity point to a general architectural principle: The parts of an organism that perform specific tasks or functions are grouped into modules that can function semi-autonomously.

The prevalence of modularity (both spatial and topological) in living systems might have several ultimate evolutionary origins (see Ref. [11] for a recent review). One long-standing idea is that modularity facilitates adaptation, in particular by enhancing the frequency at which new and useful traits appear, and by increasing their heritability. Indeed, modules that are semi-autonomous entities can be easily modified, added, replaced, or rearranged in a system through a process that has been called evolutionary tinkering [12-14]. In this view, modularity would be favored by natural selection because it modifies the *rate* of adaptation [15-17]. This scenario predicts that directional selection will bring forth organisms and networks that are highly modular; it can be particularly relevant for the evolution of complex traits [17-18]. A specific realization of this scenario arises in models of genetic network evolution when the environment is fluctuating and structured; modularity can then arise as a result of selection for a high rate of adaptation in changing environments [19-20]. But modularity in this scenario need not even require environmental change. In particular, it can emerge from innovations that allow adaptation to new ecological niches, as suggested by studies of metabolism [9-10], or from innovations that increase fitness, as suggested from gene network studies [21].

In other scenarios for the origin of modularity, natural selection on the rate of adaptation does not shape modularity; instead modular architectures follow from developmental constraints, or from other phenomena related to epistasis and pleiotropy [11]. In such scenarios, modularity can be the mere consequence of selection on other traits, but researchers do not agree on how general this scenario could be [22].

In the present work we focus on metabolism, and show that modularity in genome-scale metabolic networks may be a by-product of phenotypic constraints. We will show that this scenario is likely to be very general in metabolism for traits related to an organism's ability to live in different environments. We refer to this ability as an organism's metabolic *versatility*, and explain it further below. Specifically, we view metabolism as a complex chemical reaction network inside an organism and ask this question: Among all possible metabolic reaction networks with high versatility, do most have modular architectures?

In contrast to many other networks [23-24], metabolic networks do not just have a static graph structure, but a function that involves the flow of molecules through them. This function can be used to define modules in a network based on *fully coupled sets* of reactions, as explained in the Methods and Results sections [25-29]. We will measure a network's modularity by several indices based on these modules, and relate this modularity to versatility, a metabolism's ability to sustain life in different environments. Some organisms are metabolic specialists and can live in few environments, others are generalists that can thrive in many different environments. General principles of how a metabolic network must change as an organism's versatility varies remain elusive. One might try to find such principles by studying a broad panel of living organisms that differ in versatility. However, any association between versatility and some other observable quantity, such as modularity, would leave open whether the association between the two is driven by evolutionary forces that act not on versatility but on some other, unknown network property.

To avoid this difficulty, we can take advantage of our ability to create random samples of metabolic genotypes with specific properties, including versatility. (More precisely, the genotypes we consider are discretized binary metabolic genotypes, representations of genotypes that are suitably simplified for our purpose, as explained in Methods.) This approach [30] allows us to examine the consequences of versatility for network modularity, in the absence of any other influences. We shall find that the ability to thrive in increasing numbers of environments is strongly associated with greater modularity of metabolic networks. Our observations support the idea that elementary properties of metabolic networks, such as their ability to sustain life in multiple environments, can contribute to shaping metabolic network structure and in particular modularity, without the need to consider evolutionary dynamics or selection on a rate of adaptation.

# Modeling framework

For our study, we use genome-scale metabolic network modeling. The set of chemical reactions that can take place in an organism and their associated metabolites define the organism's metabolic network. Each reaction is typically catalyzed by an enzyme that allows the transformation of substrate molecules into product molecules. With the advent of genome-scale metabolic network modeling [31-34], it has become possible to compute which target products can be synthesized by a given set of enzymes, assuming that the network is in a metabolic steady state, and that specific nutrients are available to the network from the environment. The relevant computational method is based on balancing metabolic flux – the rate at which a reaction converts substrates into products – for all reactions, and is thus called flux balance analysis (FBA) [32, 35]. Its predictions are usually in good agreement with experimental data [36-38], except where enzymes are mis-regulated, such that a network cannot attain optimal metabolic fluxes through all its reactions. (Such mis-regulation can be sometimes eliminated during laboratory evolution experiments [37, 39], if growth rate maximization is the sole objective of the experiment.)

An organism's set of enzyme coding genes, identified here with a list of reactions, can be viewed as a discretized binary metabolic genotype; for brevity we refer to it from here on as the organism's genotype or metabolic genotype. Specifically, given a total universe of $N$ possible reactions, any genotype can be represented by a string of $N$ bits, $G=(b_1, b_2, …, b_N)$ as illustrated in Additional File 1, Figure S1a. If enzyme $i$ is encoded in the organism's genome, then $b_i=1$, while $b_i=0$ otherwise. In this framework, the space of all metabolic genotypes contains $2^N$ elements. Following previous work [30, 40], we here take the universe of reactions to encompass all known enzyme-catalyzed chemical reactions, as represented in publicly available databases [41-42]. This set of reactions is most likely incomplete, but nevertheless sufficiently comprehensive to produce a vast space of metabolic genotypes, where each genotype contains a subset of these reactions.

If an organism can grow in a specific chemical environment (defined through the nutrients it contains), its metabolic network is able to produce all of its biomass precursors (see Methods); we then call the organism (and by extension its metabolic network) *viable*. This leads us to define an organism's phenotype via its ability to synthesize biomass in a number of given chemical environments. Note that the mapping from genotype to phenotype in our approach is completely determined by the FBA framework. Previous research has shown that this map is highly degenerate, meaning that a huge number of genotypes will produce the same phenotype; indeed, many reactions in a metabolic network are typically non-essential and can be replaced by other reactions. Furthermore, genotypes of identical phenotype are such that small genotypic changes (a reaction deletion, addition, or exchange) connect these genotypes into a vast graph; we refer to this graph as a *genotype network*. A consequence of this connectivity property is that gradual evolution of genotypes is possible, while leaving the phenotype unchanged [30, 40]. For this reason, genotype networks can facilitate evolutionary changes and adaptation of genotypes [43]. Such properties seem to be generic properties of well-studied genotype to phenotype maps, and have been found in many systems. These include RNA and proteins, where the genotype is the sequence and the phenotype is the secondary or tertiary structure [44-46], as well as gene regulatory networks whose genotype specifies a pattern of genetic interactions and whose phenotype corresponds to a gene expression pattern [47].

To characterize metabolic networks of a given phenotype, we cannot examine all genotypes because of their astronomical number. Instead, we use a Markov Chain Monte Carlo (MCMC) [48] approach to sample a space of genotypes or subsets thereof. This approach is based on performing random walks within that subspace, as illustrated in Additional File 1, Figure S1c. At each step of such a random walk, a small change is applied to the current genotype and the phenotype of the changed genotype is computed; if the phenotype fulfills a pre-specified criterion, the current genotype is updated; if not, the change is rejected, and the current genotype is kept. With appropriate precautions [30] this procedure will create uniform samples of the accessible space of genotypes with a desired phenotype.

## Results

**Fully Coupled Sets of reactions are proxies of pathways**

The analysis of modularity in large graphs or networks is a mature field. Not surprisingly, multiple different measures of modularity have been developed [8, 49-54]. Identifying all modules of a large network can be computationally intractable, that is, NP difficult [55-56]. Fortunately, metabolic networks are special, because their analysis can go beyond graph-based representations. The reason is that metabolic networks synthesize biomass, and this function of metabolic networks can be quantified by studying the flow or *flux* of matter through each reaction in a network. Doing so permits an analysis of modularity that is based on network function, not just topology. Here we take advantage of the notion of *coupling* between reaction fluxes to identify sets of reactions that form a metabolic module. Such sets have been referred to as *reaction/enzyme subsets* or *correlated reaction sets* or Fully Coupled Sets [25-29]. Hereafter we will use the term Fully Coupled Set (FCS) only.  These sets define metabolic network modules that are both biochemically sensible [28-29, 57-58] and computationally tractable [28]. By definition, two reactions are in the same FCS if the ratio of their fluxes is fixed when considering all possible steady-state flux distributions through the metabolic network. Determining all FCSs of a large metabolic network can be done efficiently using linear optimization (see Methods). We note that the different FCSs in a metabolic network are disjoint, and that not all of a network's reactions need to belong to an FCS (see Methods).

The simplest possible FCS involves reactions in a linear biochemical pathway, arguably the most intuitive form of a functional module in biochemistry. However pathways with branches and cycles can also form FCSs [28]. For illustration, Figure 1 represents the largest FCS containing a cycle that arises in the *E. coli* metabolic network; it includes reactions that are involved in cell envelope biosynthesis.

We first asked how modules, as defined by FCSs, relate to conventional biochemical pathways, the classical functional modules of metabolism. To this end, we mapped reactions in many different FCSs onto biochemical pathways, as defined by standardized annotations [41, 59]. We relied on annotations in the Kyoto Encyclopedia of Genes and Genomes database (KEGG) [41], a comprehensive metabolic database that annotates biochemical reactions as belonging to a list of pathways. For the metabolic network of *E. coli*, we find that reactions in the same FCS typically belong to a common pathway.  To quantify whether this property was statistically significant, we implemented the following test.

For each FCS, we identified the pathway annotation for all of its reactions. Because each reaction can be annotated as belonging to multiple pathways, we identified for each FCS the pathway annotation that is shared by most of its reactions. We defined the quantity Q as the fraction of reactions that are annotated as belonging to that pathway, and computed Q for each FCS in the metabolic network of *E. coli*. We observed that in most FCSs (50 percent, corresponding to 50 of 100 FCSs in *E. coli*) all reactions belonged to the same pathway, and nearly 75 percent of FCSs had more than 75 percent of their reactions belonging in the same pathway. This strong association of reactions in an FCS with one pathway is not expected by chance alone, as a randomization test shows (P<0.001). Thus, most of the FCSs in *E. coli* can be viewed as biochemical pathways or parts thereof.

The same analysis can be applied to random samples of metabolic networks with specific properties, as generated by our MCMC sampling procedure (see Methods). Specifically, we first identified FCSs from 1000 *in silico* metabolic genotypes viable in all of the 89 carbon source environments we consider (see Methods). In this analysis, we observed that in most FCSs (74 percent, corresponding to 45,893 of 62,148 FCSs examined) all reactions belonged to the same pathway, and nearly 80 percent of FCSs had more than 80 percent of their reactions in the same pathway (see Additional File 1, Figure S2). Just like *E. coli*, the strong association is not expected by chance alone, as a randomization test shows (P<0.001; see Methods). Thus, both for *E. coli* and for our random samples, most FCSs can be viewed as biochemical pathways or as parts thereof. To illustrate such FCSs, Additional File 1, Figure S3 shows the most frequent FCS comprising five or more reactions that we found in our sampling. This FCS occurred in 898 of the 1000 metabolic genotypes. All its reactions belong to histidine metabolism (Q=1).

## Both measures of modularity *M* and *s* increase with versatility

We next asked quantitatively how network modularity is affected by environmental versatility. To answer this question, we defined two indices of network modularity, which we call *M* and s. The first index is the number *M* of reactions in a network that belong to FCSs. Then we calculate the average <*M*> for a sample of networks generated by our MCMC procedure, where each network in the sample needs to be viable in a given set $V_{env}$ of chemical environments (see Methods). We consider $V_{env}$ as an index of *environmental versatility* for these metabolic networks. In our analysis, we study up to 89 minimal chemical environments that differ only in the *sole* carbon source they contain (see Methods and Additional File 2, Table S1). In other words, $V_{env}$ indicates the number of sole carbon sources from which these networks can synthesize all essential biomass precursors. To see how our observations depended on the sets of carbon sources used, we investigated different choices for these sets, where sets of fewer carbon sources were nested within sets of more carbon sources (see Methods for further details).

In Figure 2a we show how <*M*> depends on versatility $V_{env}$, both on average (yellow dots), and for multiple different nested sets of carbon sources (symbols with different colors and shapes). The analysis is based on metabolic networks with the same number of reactions as *E. coli* [42]. The data show that greater versatility leads to higher values of the modularity index; this trend is clear when considering the average over all choices of carbon sources, and also when considering the different nested sets.

As a network's versatility rises, does an increase in *M* – the number of reactions in FCSs – occur through an increase in the size of the FCSs, or through an increase in the number of FCSs, while their size remains constant? To address this question, we next studied the number of FCSs, which we denote by *s*, our second index of modularity. We applied the procedures we described earlier to the same genotypes as before, averaging now the number of modules (FCSs) rather than the number of reactions in these modules.

Figure 2b shows the average number of modules, which we denote as $<s>$, for the same 1000 metabolic genotypes, the same choices of $V_{env}$, and the same nested sets of environments as above. The figure shows that greater versatility leads to higher values of this modularity index. This holds for the averages over different nested sets (yellow dots), and also without averaging, *i.e.*, for different nested sets of carbon sources (symbols of different shapes and colors). Both $<M>$ and $<s>$ show a monotonic increase with $V_{env}$ but with possible deviations from linearity.

The results of Figure 2 were obtained from networks whose number *n* of reactions equaled that of the *E. coli* metabolic network, *i.e.*, *n*=831 [42]. Additional File 1, Figures S4 and S5 show that the patterns we see are not sensitive to the number of reactions in a network. Specifically, Additional File 1, Figure S4 shows that the average number of reactions in FCSs, $<M>$, increases with versatility $V_{env}$ also for networks with *n*=500 (Figure S4a) and *n*=700 (Figure S4b) reactions. The sole difference to the data of Figure 2a is that the increase of $<M>$ is beginning to level off as $V_{env}$ reaches the largest values investigated here, in particular for *n*=500. Additional File 1, Figure S5 shows that the average number of modules, $<s>$, also increases with versatility at *n*=500 (Figure S5a) and *n*=700 (Figure S5b) reactions. However, in contrast to the trend for $<M>$ in Additional File 1, Figure S4, the increase in $<s>$ does not slow down for the largest values of $V_{env}$ we have examined.

## Modular architecture of the *E. coli* metabolic network

So far we have shown averages of our modularity measures *M* and *s* based on samples of random networks of a given versatility. In such a sample, modularity has a distribution, where some networks are more modular, and others less so. We can use this distribution to ask whether the modularity observed in the metabolic network of an organism such as *E. coli* is atypically high or low. In other words, the distribution of modularity arising in our samples of *in silico* metabolic networks can provide a null hypothesis to evaluate whether a biological network shows unusual modularity.

Figure 3a shows the distribution of the total number *M* of reactions in modules, and Figure 3b shows the distribution of the number *s* of modules (FCSs) in a sample of 1000 random networks with *n*=831 reactions (the same as *E.coli* [42]), where each network is able to sustain growth on the 89 different sole carbon sources as given in Ref. [29]. This phenotype constitutes the *in silico E. coli* metabolic phenotype we use. The figure also shows the values of *M* and *s* for the metabolic network of *E. coli*. The data from the network sample allows us to test the null hypothesis that *M* or *s* for *E. coli* could have been drawn from the sample. We find that *M* is atypically large, being in the top 3 percentile of our MCMC sample. This allows us to reject the null hypothesis at a P-value of P=0.028. Based on this analysis, we conclude that the metabolic network of *E. coli* is more modular than expected.

The architecture of the *E. coli* metabolic network has higher modularity than anticipated, but the large value of *M* may come from either a greater number of FCSs or from an increased size of the FCSs. Figure 3b shows that the number of FCSs in *E. coli* is just slightly above the position of the distribution's peak in our ensemble, well within one standard deviation. From this observation one can conclude that the atypically high modularity of the *E. coli* network stems from the fact that *E. coli* has larger modules (FCSs) but not much more modules than typical networks allowing growth on 89 carbon sources.

**Reactions in versatility-dependent FCSs are just downstream of nutrients**

Thus far, we saw that metabolic networks sustaining growth on more nutrients have higher modularity, that is, more reactions contained in modules and more modules (FCSs) (see Figure 2). We surmised that these additional reactions would be closely linked to the additional nutrients that metabolic networks must utilize as their versatility increases. In other words, these reactions and the modules they reside in presumably are needed to metabolize these nutrients, and may thus occur just downstream of them. To inquire whether this is the case, we compared the FCSs of genotypes with maximal versatility ($V_{env}$ =89) to FCSs of genotypes with $V_{env}$ =1. Specifically, we first extracted the reactions that belonged to FCSs and that occurred in more than 50 percent of the genotypes in each of the two samples. Call these sets of reactions $R_{89}$ and $R_1$, for the ensembles with $V_{env}$ =89 and $V_{env}$ =1, respectively. At a qualitative level, we find that about 90% of reactions in $R_1$ also belong to $R_{89}$. We then examined the reactions that belong to $R_{89}$ but that are not part of $R_1$, and called this set of reactions $R_{89}\backslash R_1$. Are the reactions in $R_{89}\backslash R_1$ immediately downstream of the nutrients? The notion of downstream can be made quantitative through the *Scope* algorithm [60-61]. A reaction of scope distance one can use the nutrients as its only substrates, a reaction of scope distance two can use products of reactions at scope distance at most one, and so on. (See Methods for a more detailed explanation of this scope distance.) We applied this algorithm to compare the scope distances of reactions in $R_{89}\backslash R_1$ to the scope distances of all reactions in our universe of reactions. Figure 4 shows a distribution of these distances for both groups of reactions. It indicates that reactions associated in $R_{89}\backslash R_1$ generally have smaller scope distance than other reactions. A statistical test (see Methods) shows that this difference is significant with a p-value of $10^{-5}$. In sum, reactions of modules involved in increased versatility tend to be more closely downstream of nutrients, suggesting that they typically belong to pathways metabolizing such nutrients. To illustrate this property with concrete examples, we determined which FCSs in $R_{89}\backslash R_1$ involved any of the 24 reactions occurring at scope distance 1 in Figure 4b. These FCSs have various sizes that range from 2 to 4 reactions. In Additional File 1, Figure S6 we show the three largest of these FCSs, all of them with 4 reactions, together with the pathways they belong to. These FCSs are linear pathways containing reactions of scope distance 1, 2, 3 and 4; they metabolize the nutrients fucose, rhamnose and 3-hydroxycinnamic acid.

# Discussion and conclusions

Our work took advantage of a new computational method [30, 40] that uses a combination of flux balance analysis and Markov Chain Monte Carlo sampling to create large and random samples of metabolic networks with desired properties from the space of all possible metabolic networks. The property we focused on was environmental versatility, the number of chemical environments a metabolic network can sustain life in. We studied how versatility relates to a network's modularity. For our purpose, we defined modularity as the total number of reactions contained in fully coupled sets. We found that more versatile networks are more modular (they have more modules and more reactions contained in modules) than less versatile networks. We emphasize that this does not result from the fact that networks with more reactions are more versatile, because our analysis uses networks with fixed number of reactions. The reactions that form part of newly arising modules in highly versatile networks tend to be close to reactions that process nutrients. The advantage of using random samples of metabolic networks with a specific property for our analysis is that such samples have not been subject to any of the (usually unknown) selection pressures that an organism's metabolism is subject to, and that they can form a useful reference point to ask whether any one organism's metabolic network has typical or atypical properties. In such a comparison, we learned that *E. coli*'s network is significantly more modular than random networks of the same versatility, a feature arising mainly from the fact that it contains larger modules.

    Modularity in metabolic networks has been studied by several other authors [6-10]. Metabolic networks can be represented as graphs, allowing one to study topological (graph-based) measures of modularity; this approach has been taken for metabolic and other systems, such as protein interaction networks. Unfortunately, for any sensible definition of modularity, graph-based module identification is typically computationally very expensive, so in practice one resorts to heuristic algorithms to extract modules [49, 51-52]. Additionally, in graph-based representations of metabolism, many metabolites have very high degree (number of reactions they participate in). This feature may prevent any clear modules from arising, although various heuristic tricks, such as removing high degree metabolites [6-10] can be used to skirt this problem. Problems like these can be avoided by using *functional* measures of modularity. Commonly used measures involve elementary flux mode or extreme pathways [25-27], but they are ill-suited for genome-scale modeling because of the complexity in computing them. The measure of modularity we used here was based on the reactions contained in fully coupled sets (FCSs) [28]. We showed that most or all reactions in a fully coupled set fall within a single metabolic pathway, which underlines the biochemical relevance of our definition of modularity. Two further technical advantages come with our definition of modularity based on FCSs: (1) the approach involves no adjustable parameters; (2) identification of FCSs is computationally efficient even for genome-scale networks.

    Intriguingly, the extent of modularity found in *E. coli* is higher than in our *in silico* genomes. *E. coli* both has more fully coupled sets and larger fully coupled sets than expected for networks with the highest versatility we consider. This high modularity may reflect the fact that *E. coli* is even more versatile than the most versatile networks in our samples, networks that are viable on 89 carbon sources. For example, it can also grow on sources of sulfur or nitrogen that we did not consider. The high computational cost of our analysis in multiple environments currently prohibits us from extending our study to a larger spectrum of environments. Conversely, the high modularity of *E. coli* might also be caused by other factors, for example, a long record of past evolutionary adaptations that may favor modularity through the high rates of adaptation it may allow and/or its high heritability, e.g. through horizontal gene transfer [62-63]. Indeed it has been shown that FCSs and operons in *E. coli* are positively associated [29, 57-58]. Only future work will be able to validate which of these causes is more important in *E.coli*. Our network sampling approach has the advantage that it provides a rational expectation

for how modular a network can be expected to be based solely on phenotypic constraints. It thus puts answering this question within reach.

Given the ubiquity of modularity in biological systems, it is tempting to propose general principles that might explain its appearance. By comparing natural with man-made systems and following the original insights of Jacob [12] and others [1-6, 8-11], it seems very plausible that modularity should emerge during adaptive evolutionary trajectories because it can increase the rate of adaptive change. This holds true in particular in artificial systems such as factories, companies and even industries, where modularity allows for lower costs and enhanced possibilities for innovation [64-66]. As long as a lineage of organisms is experiencing adaptive evolutionary change, modularity should remain ubiquitous, whereas in long periods of stasis modularity may become reduced. This perspective is appealing but other factors may also influence modularity, which can be seen by considering the modularity of eukaryotic cells. The organization of cells into parts with specialized tasks (organelles, ribosomes, etc.) suggests that cellular tasks are best performed in specialized modules. One may thus conjecture that modularity has not only the indirect benefit of accelerating the rate of evolutionary change, but also direct benefits such as the possibility to perform certain tasks better, and thereby allow organisms to be better adapted to the complex world around them.

The question whether biological modularity may have a direct benefit can be addressed in systems where a realistically complex yet computationally tractable genotype to phenotype relationship exists. Genome-scale metabolic network models are such systems [32]. Answering the question amounts to finding out whether the best performing genotypes (according to some criterion) have a modular architecture. The criterion we used is based on the complex trait we called environmental versatility, the number of environments a metabolic network can sustain life in. The answer we found is clear: Requiring viability in additional environments requires additional pathways or modules to metabolize more nutrients and thus versatility enhances modularity.

Our analysis shows that modularity can be a by-product of versatility, at least in the framework of our metabolic modeling, because our system has no selective pressure on modularity per-se; highly versatile networks that are also highly modular are simply more numerous than the less modular ones. In the language of constraint satisfaction problems [67], constraints are easier to satisfy using modular architectures, so highly modular solutions will be more numerous than the less modular ones. An analogy with the engineering of network architectures may be appropriate here. Consider the circuit layout problem where a circuit's electronic components and wires must be placed on a chip. If no constraints are imposed on the circuit's speed, many different layouts are possible. But, if one focuses on the fastest circuits, one will find that they have shorter wires and are more modular, so modularity is a by-product of circuit speed. In this example, functional constraints change the architectural characteristics in the space of possible solutions. Such a property may be expected to arise in both artificial and natural systems.

Since versatility corresponds to viability in increasing numbers of environments, it can be considered as a trait associated with fitness itself. Our work suggests that modularity can emerge as a consequence of increasing functional constraints. Because our work is not just based on one or few metabolic networks from well-studied organism, but on large samples of random viable networks, we also suggest that this scenario may be generally important. Recent observations by Parter et al. [9] and Kreimer et al. [10] where generalists prokaryotes living in many different environments are more modular than specialists are fully consistent with this conclusion.

# Methods

### Flux Balance Analysis (FBA)

Flux balance analysis (FBA) [32, 35] is a computational modeling approach widely used to analyze genome-scale metabolic networks. FBA uses structural information contained in the stoichiometric coefficients of each reaction in a metabolic network to predict the possible steady state fluxes of all reactions and the maximum biomass yield of an organism. FBA does not require knowledge of metabolite concentrations or detailed information of enzyme kinetics. The stoichiometric coefficients of all reactions in a network are encapsulated in a matrix **S** of dimensions $m$ x $n$, where $m$ is the number of metabolites and $n$ is the number of reactions. Note that some of these reactions correspond to transport processes, *i.e.*, they import or export metabolites. In a metabolic steady state, such as might be attained in a growing cell population with adequate nutrient supply, the metabolites achieve a dynamic mass balance wherein the vector **v** of metabolic fluxes through the reactions satisfies the equation

$$\mathbf{Sv} = 0 \qquad (1)$$

so as to satisfy mass conservation. Eq. 1 represents stoichiometric and mass balance constraints on the metabolic network. For genome-scale metabolic networks, Eq. 1 leads to an under-determined system of linear equations in the metabolic fluxes, leading to a large solution space of allowable fluxes. The space of allowable solutions can be reduced by incorporating thermodynamic constraints associated with irreversible reactions, as well as flux capacity constraints which limit the maximum flux through certain reactions. To obtain a particular solution, linear programming (LP) is used to find a set of flux values – a point in the solution space – that maximizes a biologically relevant linear objective function $Z$. The LP formulation of the FBA problem can be written as:

$$\max Z = \max \{\mathbf{c}^T \mathbf{v} \,|\, \mathbf{Sv} = 0, \mathbf{a} \leq \mathbf{v} \leq \mathbf{b}\} \qquad (2)$$

where the vector **c** corresponds to the coefficients of the objective function $Z$, and vectors **a** and **b** contain the lower and upper limits of different metabolic fluxes in **v**. The objective function $Z$ is often chosen to be the so-called growth flux. This is the flux through an artifactual reaction that reflects the synthesis of biomass, which requires biosynthesis of biomass precursor molecules, such as amino acids and nucleotides, in specific proportions. The stoichiometry of this reaction is based on the experimentally measured biomass composition of an organism. The predictions from the FBA framework and related approaches are often in good agreement with experimental results [36-38, 68].

### Reaction database

In this work, we have used a hybrid database compiled by Rodrigues and Wagner [40] containing 4816 metabolites and 5870 reactions. This hybrid database was obtained by merging the reactions in the Kyoto Encyclopedia of Genes and Genomes (KEGG) LIGAND [41] with those in the *E. coli* metabolic model iJR904 [42], followed by appropriate pruning to exclude generalized polymerization reactions. Of the 5870 reactions, 2501 are reversible and 3369 are irreversible. Note that more than 5500 reactions in the hybrid database are contained in KEGG database and less than 300 reactions are specific to the *E. coli* metabolic model iJR904.

In addition to the 5870 metabolic reactions, the hybrid database has transport reactions for 143 external metabolites contained in the *E. coli* iJR904 model. These 143 external metabolites were assumed to be the set of possible imported and secreted metabolites. Further, the hybrid database includes an objective function $Z$ in the form of a biomass reaction that reflects synthesis of the *E. coli* biomass components, as defined in the iJR904 model.

Genome-scale metabolic networks typically contain *blocked* reactions [28, 69] which are dead-ends and necessarily have zero flux for every examined chemical environment under any steady-state condition. Such blocked reactions cannot contribute to any steady-state flux distribution and can be excluded from the hybrid database. For the set of 143 external metabolites, we found 2968 of the 5870 reactions in the hybrid database to be blocked under all environmental conditions we examined. We have excluded this set of 2968 blocked reactions from the hybrid database of 5870 reactions to arrive at a reduced reaction set of 1597 metabolites and 2902 reactions. We thus take this reduced set of $N$=2902 reactions as the *global reaction set*.

The *E. coli* metabolic model iJR904 has 931 reactions (which of course are contained in the hybrid database of 5870 reactions). Our global reaction set (having 2902 reactions) was derived from the hybrid database by excluding blocked reactions; after this exclusion, the *E. coli* metabolic model iJR904 is left with 831 reactions. Here, we consider this set of 831 reactions to be the *E. coli* metabolic genotype.

**Viable genotypes**

Any subset of *n* reactions taken from the global reaction set is considered to specify a discretized binary metabolic genotype. For simplicity, we shall refer to this as a metabolic genotype or as a genotype. Specifically, a metabolic genotype $G$ can be represented by a bit string of length $N$, i.e., $G=(b_1, b_2, …, b_N)$, where $N$ is the number of reactions in the global reaction set (see Additional File 1, Figure S1a). Each position in the bit string $G$ corresponds to one reaction in the global reaction set, with the reaction being either present ($b_i$=1) or absent ($b_i$=0) in the genotype. We denote the set or space of metabolic genotypes with a given number *n* of reactions as $\Omega(n)$.

For any genotype, we can use FBA to determine whether the corresponding metabolic network has the ability to synthesize all biomass components in a given chemical environment (medium). We consider a genotype to be *viable* in a given environment if and only if the maximum biomass flux predicted by FBA for the genotype is nonzero; otherwise we consider the genotype to be non-viable (see Additional File 1, Figure S1b). In general, *in silico* metabolic studies take a metabolic network's *fitness* to be proportional to the maximum biomass growth flux the network can attain in a given environment. The metabolic property considered here is simpler: we ask only whether a network can synthesize all biomass components in a given environment, regardless of the synthesis rate. For all the work we report, we use the *E. coli* biomass composition to determine the viability of a genotype in a given chemical environment.

## Chemical environments and phenotypes

For our purpose, the metabolic phenotype of a metabolic network (genotype) is determined by the network's viability in a list of well-defined chemical environments (media). We shall denote the subset of genotypes within $\Omega(n)$ that have a specific phenotype – growth on a specific list of environments – as $V(n)$. In this work, we use only aerobic minimal environments containing one carbon source. Each environment also contains unlimited amounts of the following inorganic metabolites: ammonia, iron, potassium, protons, pyrophosphate, sodium, sulfate, water and oxygen. Based on FBA applied to the metabolic model iJR904, it was found earlier that *E. coli* can support nonzero biomass growth on 89 different aerobic minimal environments [29, 57]. The environments we focus on here differ in these 89 carbon sources, which are listed in Additional File 2, Table S1.

## Environmental versatility index ($V_{env}$) and nested choices of chemical environments

The Markov Chain Monte Carlo (MCMC) sampling algorithm (see also below) can be used to explore the set of genotypes having a given phenotype. In our case, this phenotype is viability on a given set of minimal environments; if this set consists of $V_{env}$ environments, we say that the genotype's environmental versatility index is $V_{env}$. Thus, the phenotype $V_{env}=1$ refers to genotypes viable in one specific environment, the phenotype $V_{env}=2$ refers to genotypes viable in two given environments, and so on. We have considered 89 minimal environments whose sole carbon sources, their only distinguishing feature, are listed in Additional File 2, Table S1.

     We have used MCMC to sample ensembles of increasingly versatile metabolic networks, *i.e.*, ensembles whose networks have $V_{env}=1, 2, 5, 10, 20, 30, 40, 50, 70$ and 89. The genotypes with $V_{env}=89$ are the most versatile among them as they are viable in all 89 minimal environments. There are many ways of choosing 1, 2, or more specific environments out of 89 environments to sample genotypes having a phenotype with $V_{env}=1, 2$, through 89. The properties of sampled genotypes in an ensemble with a given $V_{env}$ will depend on the choice of those $V_{env}$ environments. The computations we carry out are computationally very expensive, and they become more expensive with every additional environment in which viability is determined. To limit this expense, we pursued two strategies. First, we used *nested* sets of environments to sample genotypes in ensembles with different $V_{env}$, e.g., the set of 70 environments chosen for $V_{env}=70$ is a subset of that for $V_{env}=89$, and the set of 50 environments chosen for $V_{env}=50$ is a subset of that used for $V_{env}=70$, and so forth. Second, we used only one subset of 70 environments within $V_{env}=89$ to sample an ensemble with $V_{env}=70$, and only one subset of 50 environments within the choice for $V_{env}=70$ for sampling an ensemble with $V_{env}=50$. For $V_{env}$ below that, we did tackle the variability coming from different environmental choices; specifically, for $V_{env}=40$, we used 10 different subsets of 40 environments within the choice for $V_{env}=50$; thus we generated 10 different genotype ensembles, where each genotype in each ensemble had $V_{env}=40$. Each of these 10 different choices of 40 environments was then used to create a single nested sequence for $V_{env}=30, 20, 10, 5, 2,$ and 1. This allowed us to have 10 different ensembles to sample at each of these $V_{env}$ and to follow for each sequence of nested sets the consequences of modifying $V_{env}$. (See Additional File 1, Figure S7 for a diagram representing two such nested sets.) We

computed the average properties of the sampled genotypes as well as their dispersion based on the 10 different samples for each value of $V_{env}$.

## MCMC sampling of viable genotypes

It was shown in previous work [30] for a single environment, corresponding to $V_{env} = 1$, that the size of the subspace $V(n)$ relative to $\Omega(n)$ is of the order of $10^{-22}$ for genotypes with $n = 2000$ reactions. This size decreases even further if one requires viability in multiple environments. Such tiny probabilities of finding a desired phenotype in $\Omega(n)$ make it infeasible to sample genotypes in $V(n)$ by simply drawing random genotypes in $\Omega(n)$ with the correct number $n$ of reactions, followed by determining the phenotype of each genotype. Thus, we relied on the Markov Chain Monte Carlo (MCMC) method described in Ref. [30] to uniformly sample genotypes in $V(n)$.

This MCMC method starts with a genotype in $V(n)$ and produces a sequence of genotypes, wherein the $(k+1)^{th}$ genotype in the sequence is generated from the $k^{th}$ genotype using a probabilistic transition rule. At each transition step, one proposes a small modification to the current genotype in the sequence; if the modified genotype has the correct phenotype, one accepts the modified genotype as the next genotype of the sequence; otherwise the next genotype becomes identical to the current genotype. The modification introduced at each transition step is a reaction swap. It consists of removal of one reaction from the current genotype, followed by addition of new reaction from the global reaction set to generate a modified genotype. Note that the reaction swap preserves the number $n$ of reactions in the genotype (see Additional File 1, Figure S1a). Thus, the MCMC approach produces a walk in the subspace $V(n)$, as illustrated in Additional File 1, Figure S1c. Note that in the limit of long walks, this approach samples uniformly the space of genotypes that are accessible from the first genotype of the MCMC procedure and that have a given phenotype.

In our simulations, starting from an initial genotype in $V(n)$, we have first carried out $10^5$ attempted swaps to erase the memory of the starting genotype. After this initial phase, we continued the MCMC procedure to sample genotypes in $V(n)$. During this later phase, it is not useful to keep all of the genotypes produced, as many of them may be highly similar to one another. We thus saved only every $1000^{th}$ genotype generated in a sequence of $10^6$ steps. This procedure produces a random ensemble of 1000 genotypes in $V(n)$ [30]. We sampled genotypes in $V(n)$ for three different values of the number of reactions $n = 500$, 700 and 831.

To start the MCMC sampling, a first genotype having the correct phenotype is required. To this end, we first determined those reactions in the *E. coli* metabolic network that have nonzero flux in an optimal flux distribution with maximum biomass flux for each of the 89 minimal environments considered here. (Recall that *E. coli* is viable on all of our 89 environments). The number of nonzero fluxes is ~300 in a typical optimal flux distribution for each of the 89 environments. We generated a genotype with $n$ reactions and phenotype $V_{env} = 1$ (*i.e.*, growth on one specified environment) by starting with the set of nonzero flux reactions for *E. coli* in that environment, and then adding randomly other reactions until we reached a metabolic genotype with exactly $n$ reactions. We generated a genotype with $n$ reactions and $V_{env} = 2$ (*i.e.*, growth in two specific environments) by starting with the union of the two sets of reactions that had nonzero flux when the *E. coli* metabolic network synthesized biomass in the two different environments; then we added randomly other reactions until we reached a metabolic genotype with exactly $n$ reactions. We generated starting genotypes with $V_{env}$ =5, 10, 20, 30, 40, 50, 70 and 89 analogously.

## Fully coupled sets (FCSs) and measures of modularity

A reaction pair $v_1$ and $v_2$ are said to be *fully coupled* to each other if a nonzero flux for $v_1$ implies a proportionate (nonzero) flux for $v_2$ in any steady state and vice versa [28]. A *fully coupled set* (FCS) in a metabolic network is a maximal set of reactions that are mutually fully coupled to each other (thus, there are no FCSs of size 1). A simple argument shows that FCSs of a network are non-overlapping entities. Indeed, if a reaction were to belong to two FCSs, then all reactions in those two sets would be fully coupled pairwise, resulting in one larger FCS.

We denote the number of FCSs in a metabolic network genotype by $s$. This is one index of modularity. We also define the modularity index $M$ for a genotype as the number of reactions contained in the FCSs of that genotype ($M$ can vary from zero to the total number of reactions in the network). Burgard *et al* [28] have proposed a linear programming (LP) based method to determine whether two fluxes in a metabolic network are fully coupled. The LP formulation of the coupling problem can be written as:

$$\text{Solve } R_{max} = \max\{v_1/v_2 = 1, \mathbf{S}.\mathbf{v} = 0, \mathbf{a} \leq \mathbf{v} \leq \mathbf{b}\} \quad (3)$$

$$\text{Solve } R_{min} = \min\{v_1/v_2 = 1, \mathbf{S}.\mathbf{v} = 0, \mathbf{a} \leq \mathbf{v} \leq \mathbf{b}\} \quad (4)$$

If $R_{max} = R_{min}$ then $v_1$ and $v_2$ are fully coupled. In the above equations, $\mathbf{S}$ is the stoichiometric matrix, and vectors $\mathbf{a}$ and $\mathbf{b}$ contain the lower and upper limits of different metabolic fluxes in $\mathbf{v}$.

We have used the algorithm of Burgard *et al.* to determine all FCSs in our metabolic network genotypes. We have computed the coupled reaction pairs under conditions where all external metabolites were allowed to be imported or secreted. Further, coupled reaction pairs were computed without assuming a constant biomass composition to avoid coupling a large set of fluxes to the biomass reaction. Hence, all biomass components were allowed to be synthesized independently of one another, without constraining their stoichiometry in the biomass.

## Scope algorithm and distance of reactions from nutrient metabolites

Ebenhöh and colleagues [60-61] have introduced the concept of *scope* based on a network expansion algorithm for the structural analysis of genome-scale metabolic networks. Their approach calculates for a given metabolic network/reaction database and predefined external metabolites (referred to as seed metabolites) the set of metabolites – the scope – which the reaction network is in principle able to produce. In other words, the scope describes the synthesizing capacity of a given set of seed metabolites given a list of metabolic reactions.

The Scope algorithm iteratively updates a set $A$ of metabolites that a reaction system can synthesize. In the algorithm this set $A$ is initialized to the set of nutrient metabolites. At each iteration $i$ of the algorithm, one takes the current set $A(i)$ of producible metabolites and expands it to set $A(i+1)$ as follows. First one initializes, $A(i+1)$ to contain all metabolites in $A(i)$. Then one considers successively each reaction in the database and adds that reaction's products to $A(i+1)$ if and only if all of its substrates are in $A(i)$. This procedure ends when $A(i) = A(i+1)$, that is when in a given iteration no new molecules can be synthesized. We have used the Scope algorithm to define a distance of a reaction in the global reaction set from nutrient metabolites. Specifically, the distance of a reaction from the nutrient metabolites in the seed set is defined as the iteration number *i* of the Scope algorithm when that particular reaction contributes its products to *A(i+1)*.

A limitation of the Scope algorithm in comparison to constraint-based frameworks like FBA is its inability to deal properly with the self-generating (autocatalytic) nature of certain cofactor metabolites (e.g., ATP, NADH) in the network [70-71]. The scope of the nutrient metabolites in the seed set is sensitive to the presence or absence of such co-factors in the seed set. Following Kun et al [71], we included in the seed set the autocatalytic metabolites listed in Additional File 2, Table S2 (in addition to the nutrient metabolites in our minimal environment) when computing the distance of reactions with the Scope algorithm.

We have determined the distance from nutrient metabolites for each reaction in the global reaction set for 89 different seed sets corresponding to the 89 aerobic minimal environments. For each reaction, we have designated the minimum of the 89 distances obtained for the 89 different environments as the *scope distance* of that reaction from the nutrient metabolites.

## Statistical tests for the increase in modularity *M* with $V_{env}$

Since the modularity index *M* and the number *s* of FCSs is larger for sampled genotypes with $V_{env}=89$ than with $V_{env}=1$ (cf. Figure 2a), it is appropriate to identify reactions contributing to additional FCSs in genotypes having $V_{env}=89$. To this end, we combined the lists of FCSs for each of the 1000 sampled genotypes at $V_{env}=89$ (and *n*=831 reactions) to create a merged list of FCSs that occur in at least one such sampled genotype. Since the FCSs are non-overlapping entities, multiple copies of a FCS in the merged list signify the FCS's presence in multiple sampled genotypes. We then determined the set of reactions that occurred in at least 500 FCSs in the merged list of FCSs for $V_{env}=89$. We refer to it as the *consensus set* $R_{89}$ of FCS reactions for $V_{env}=89$ and *n*=831. In a similar way, we obtained the consensus set $R_1$ of FCS reactions for our sampled genotypes with $V_{env}=1$.

The consensus set $R_{89}$ for $V_{env}=89$ is larger than the set $R_1$ for $V_{env}=1$, and the complement set $R_{89}\backslash R_1$ consisting of the reactions belonging to $R_{89}$ but not $R_1$ gives the set of reactions that mostly account for the additional FCSs in $V_{env}=89$. We then considered the scope distances of reactions from nutrient metabolites for two choices of reaction sets. The first set is this complement set $R_{89}\backslash R_1$, the second is the set of all reactions in the global reaction set. (In Figure 4 we show the corresponding distributions.) The scope distances for the reactions in $R_{89}\backslash R_1$ are clearly concentrated at much smaller values than when considering *all* possible reactions. A Kolmogorov-Smirnov (K-S) test allowed us to reject the null hypothesis that the two distributions are the same ($p<10^{-5}$). Further, a two sample Welch t-test allowed us to reject the hypothesis that the means of the two distributions are the same ($p<8.10^{-8}$).

## Use of pathway classification of reactions to characterize biochemical relevance of FCSs

We have classified reactions in our global reaction set into different biochemical pathways using the pathway information [72] for reactions in the KEGG database [41], along with subsystem information [73] for the remaining reactions in the *E. coli* metabolic model iJR904 [42]. We have used this pathway classification as follows to test whether the majority of reactions in a given FCS belong to a common biochemical pathway.

For a given FCS, we define the quantity $Q$ which is the fraction of reactions sharing the dominant annotation for that FCS. We computed $Q$ for each FCS in the merged list of FCSs from our 1000 sampled genotypes with phenotype $V_{env}$=89 and $n$=831 (see the previous section for merged lists of FCSs). We then considered the cumulative distribution of $Q$ for FCSs in the merged list, namely, the probability that $Q$ is at least as large as a given value X. The cumulative distribution of $Q$ for FCSs in the merged list with $V_{env}$=89 and $n$=831 is shown in Additional File 1, Figure S2. We also computed the fraction $h$ of FCSs in the merged list with $Q \geq h$, a quantity that is analogous to the *h*-index commonly used to measure scientific productivity [74]. This *h*-index has a value of $h$=0.79, as can be seen from the point of intersection of the cumulative distribution of $Q$ with the bisecting line in Additional File 1, Figure S2.

To test the significance of the *h*-index obtained from the merged list of FCSs for sampled genotypes with $V_{env}$=89 and $n$=831, we performed the following randomization test. Starting from the merged list of FCSs and the pathway annotations of their reactions, we generated 1000 equivalent random lists by swapping the annotations among reactions in different FCSs, while preserving the frequency of each annotation in the merged list. We swapped annotations as follows. We first recorded the multiplicity of each distinct FCS within the merged list. We then randomly picked two FCSs in the merged list with the same multiplicity, and one random reaction in each of the two FCSs, and then swapped the annotations of these two reactions in the FCSs. We performed at least $10^7$ swaps starting from the merged list before saving a *random list*, that is, a list of FCSs whose reaction annotations had been randomized in this way. None of the 1000 random lists we generated had an *h*-index greater than 0.79 obtained for the merged list with $V_{env}$=89. On this basis, we can reject the hypothesis that reaction annotations are not similar within FCSs at a p-value of less than 0.001.

## Author contributions

The project was defined by all three authors. OCM and AS conceived the algorithmic procedures. AS wrote the code and performed the numerical simulations. All authors contributed in designing research, analyzing the data, and writing the paper. All authors have read and approved the manuscript.

## Acknowledgements

We thank Dominique de Vienne, Christine Dillmann and Vincent Fromion for comments, and Pierre-Yves Bourguignon for discussions. AS acknowledges support from CNRS GDRE513. AW acknowledges support through Swiss National Science Foundation grants 315200-116814, 315200-119697, and 315230-129708, as well as through the YeastX project of SystemsX.ch, and the University Priority Research Program in

Systems Biology at the University of Zurich.  OCM acknowledges support from the Agence Nationale de la Recherche, Metacoli grant ANR-08-SYSC-011. The LPTMS is an Unité de Recherche de l'Université Paris-Sud associée au CNRS.

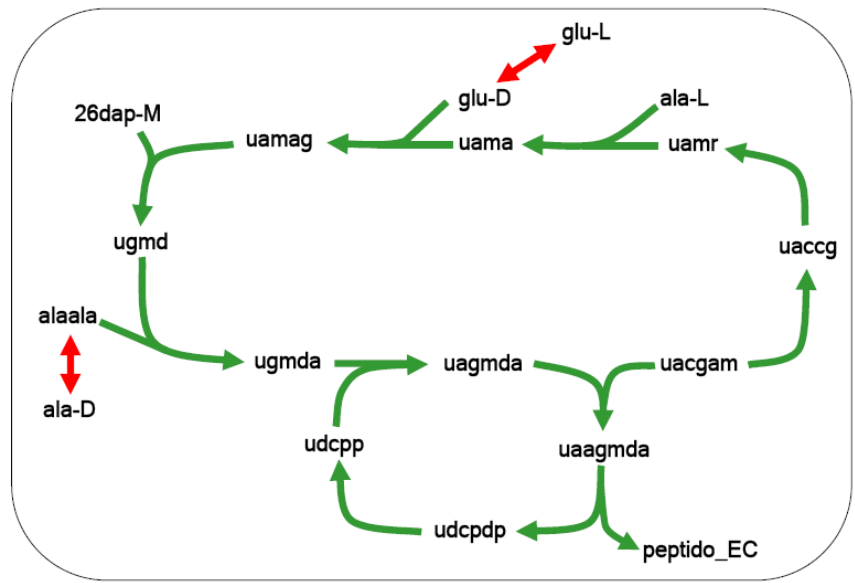

**Figure 1: Example of a FCS in the *E. coli* metabolic network.** We display a FCS of 12 reactions in the *E. coli* metabolic network that is branched and contains a cycle. In this figure, the (hyper) edges represent reactions involving metabolites. Green edges represent irreversible reactions and red edges represent reversible reactions. To reduce clutter, the ubiquitous high degree metabolites such as ATP, NADH, etc. have been omitted from this figure. Abbreviations of metabolite names are as follows: 26dap-M = meso-2,6-Diaminoheptanedioate; ala-D = D-Alanine; ala-L = L-Alanine; glu-D = D-Glutamate; glu-L = L-Glutamate; peptido_EC = Peptidoglycan subunit; uaagmda = Undecaprenyl-diphospho-N-acetylmuramoyl-(N-acetylglucosamine)-L-ala-D-glu-meso-2,6-diaminopimeloyl-D-ala-D-ala; uaccg = UDP-N-acetyl-3-O-(1-carboxyvinyl)-D-glucosamine; uacgam = UDP-N-acetyl-D-glucosamine; uagmda = Undecaprenyl-diphospho-N-acetylmuramoyl-L-alanyl-D-glutamyl-meso-2,6-diaminopimeloyl-D-alanyl-D-alanine; uama = UDP-N-acetylmuramoyl-L-alanine; uamag = UDP-N-acetylmuramoyl-L-alanyl-D-glutamate; uamr = UDP-N-acetylmuramate; udcpdp = Undecaprenyl diphosphate; udcpp = Undecaprenyl phosphate; ugmd = UDP-N-acetylmuramoyl-L-alanyl-D-gamma-glutamyl-meso-2,6-diaminopimelate; ugmda = UDP-N-acetylmuramoyl-L-alanyl-D-glutamyl-meso-2,6-diaminopimeloyl-D-alanyl-D-alanine.

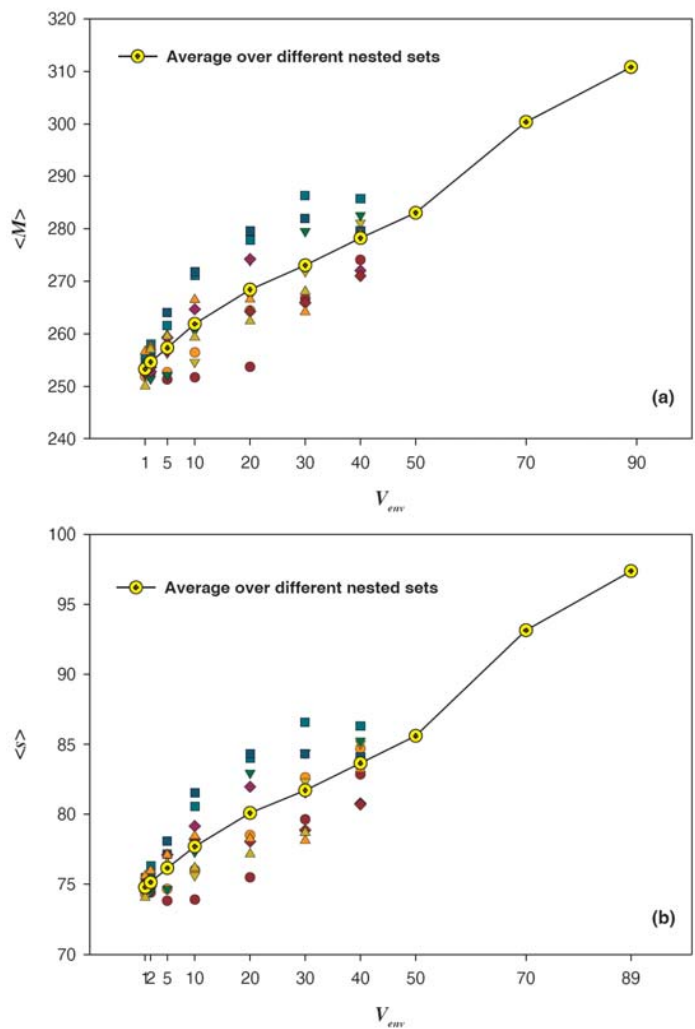

**Figure 2: A higher modularity index *M* and a greater number of modules *s* are by-products of increasing environmental versatility.** The Environmental Versatility Index ($V_{env}$, horizontal axis in both (a) and (b)) denotes the number of minimal environments in which a genotype is forced to be viable. The modularity index *M* (vertical axis in (a)) for a genotype gives the number of reactions contained in the FCSs of that genotype. The number of FCSs (modules) in a metabolic network genotype is denoted by *s* (vertical axis in (b)). The figure shows that with increasing $V_{env}$, both *M* and the number of modules *s* in a genotype increase. The data shown here are based on MCMC sampled genotypes with *n*=831 reactions (as in the *in silico E. coli* metabolic model), and 10 different choices for nested sets of environments when requiring viability on more and more environments. Each choice of nested set is displayed with a different color and symbol in (a) and (b). Each of the 10 nested sets, as well as their average (line shown for visual guidance), show a clear rise in the average of *M* (panel a) and the average of *s* (panel b) as one increases $V_{env}$.

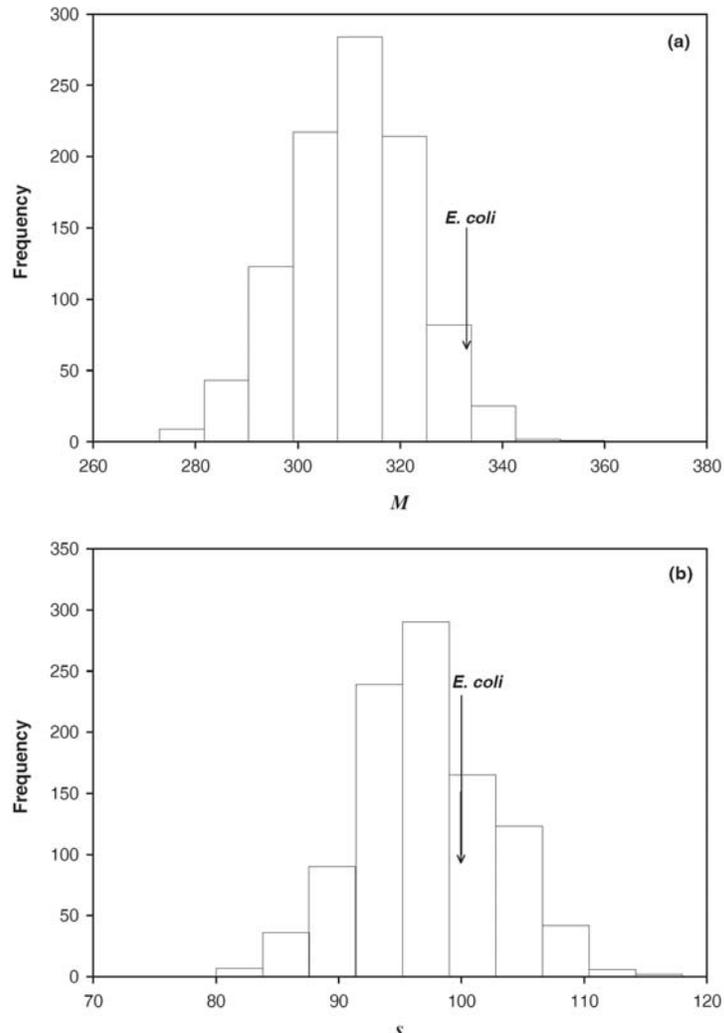

**Figure 3: Distribution of *M* and of the number of modules *s* for genotypes of phenotype with $V_{env}$ =89 in an ensemble and comparison with *E. coli*.** The horizontal axis shows the modularity index *M* in (a) and the number of modules *s* in (b). The vertical axis shows the frequency of genotypes with the corresponding value of *M* (panel a) and *s* (panel b) in a random sample of 1000 genotypes (*n*=831 reactions each, as in the *in silico E. coli* metabolic model) that are viable in $V_{env}$ =89 different minimal environments. In both panels, the histogram is displayed along with estimates of *M* and *s* for *E. coli*. From (a), we can reject at a p-value of 0.028 the hypothesis that the modularity index *M* of *E. coli* is drawn from this same distribution. Thus, *E. coli* can be considered as atypically modular.

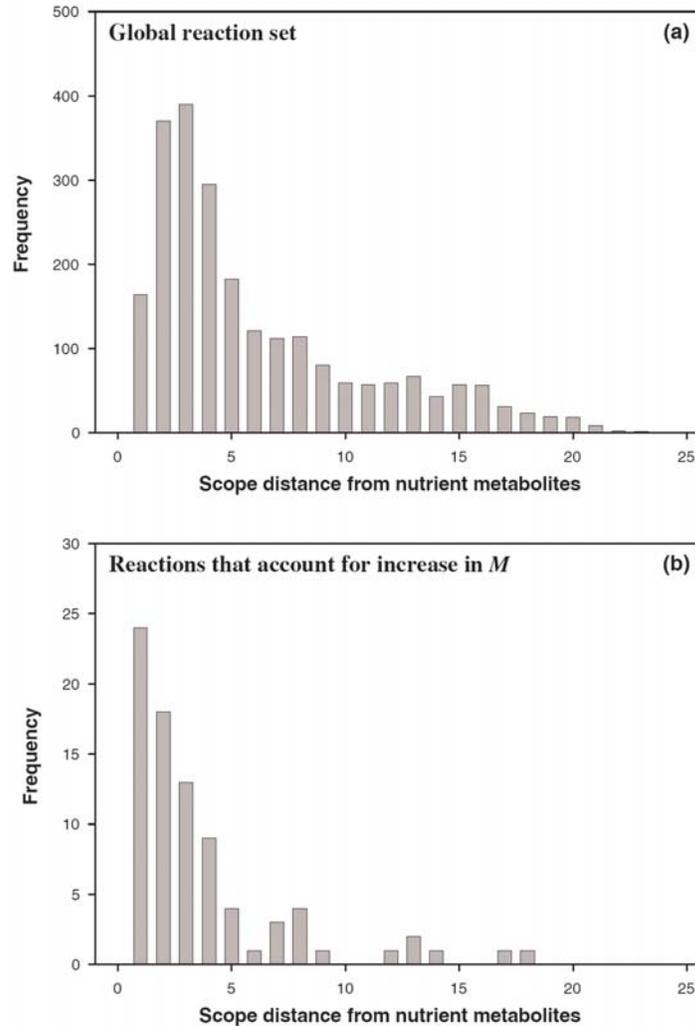

**Figure 4: The increase in modularity with $V_{env}$ can be attributed to reactions that are close to nutrients.** The sampled genotypes with $V_{env}$ =89 typically have additional FCSs compared to sampled genotypes with $V_{env}$ =1. The reactions in these additional FCSs are not typical of the whole reaction network, and instead cluster at small distances from the nutrients. (See Methods for the determination of these distances using the *Scope* algorithm.) The distribution of distances for these reactions in additional FCSs is clearly concentrated at much smaller values than the distribution for all possible reactions; a Kolmogorov-Smirnov (K-S) test yields a p-value of $10^{-5}$, allowing us to reject the hypothesis that the two distributions are the same. Furthermore, a two sample Welch t-test gives a p-value of $8\times10^{-8}$, allowing us to also reject the hypothesis that the mean of the two distributions are the same. **a)** Distribution of scope distances from the nutrients for all possible reactions. **b)** Distribution of scope distances from the nutrients for those reactions belonging to additional FCSs that differentiate the sampled genotypes at $V_{env}$ =89 from those at $V_{env}$ =1.

# Additional File 1: Figures S1 to S7

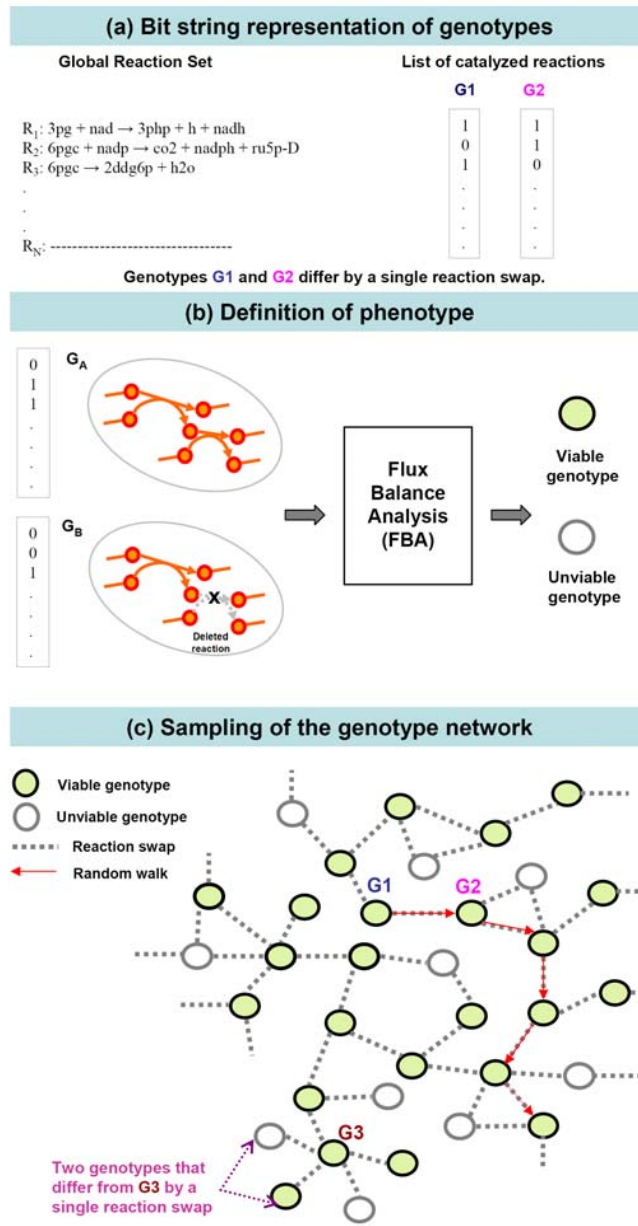

**Figure S1: The MCMC sampling of genotypes of a given phenotype. a)** A (discretized binary) metabolic network genotype refers to any subset of reactions taken from the global set of *N* reactions. Such a genotype can be represented as a bit string of length *N*; if the genotype has *n* reactions, the bit string will have *n* entries equal to 1 and all others equal to 0. **b)** For any such genotype, we can use *in silico* flux balance analysis (FBA) to determine the maximum possible biomass flux in a given environment. A genotype is considered *viable* in a given environment if and only if FBA predicts nonzero growth for that environment. A genotype has the desired phenotype if it is viable in a given list of environments. **c)** The set of all genotypes of given phenotype and having exactly *n* reactions forms a "genotype network". The MCMC algorithm samples the genotype network via a long random walk among genotypes of interest, performing at each step a reaction swap. If the new genotype has the same phenotype, the step is accepted, otherwise the step is rejected. This process is repeated many times until an adequate sampling of the genotype network is obtained.

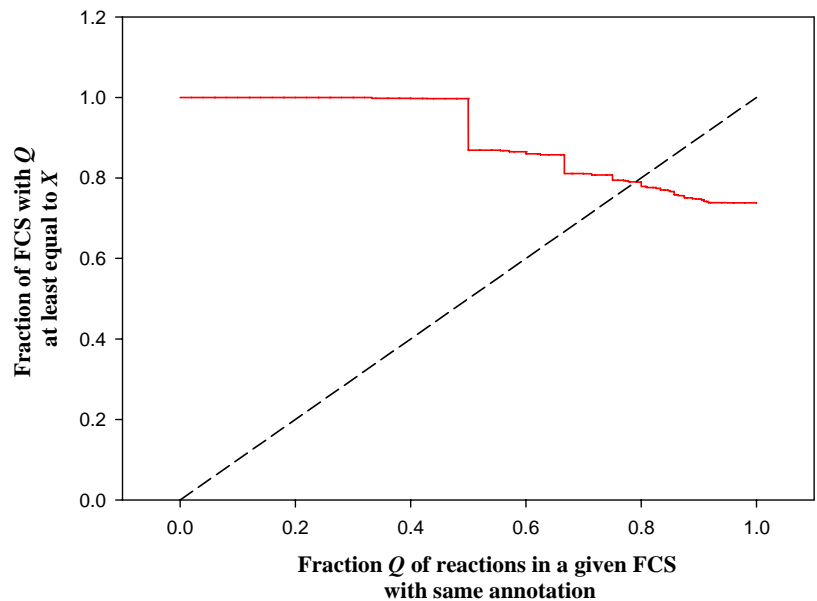

**Figure S2: A FCS predominantly consists of reactions belonging to one biochemical pathway.** For each FCS obtained using the 1000 sampled genotypes with $V_{env}$ =89 and $n$=831 reactions (as in the *in silico E. coli* metabolic model), we determined the pathway annotation of each reaction it contained. If a FCS corresponds to a biochemical pathway, its reactions should predominantly have the same annotation. To test this hypothesis we defined the quantity $Q$ for each FCS as the fraction of reactions sharing the dominant annotation for that FCS. Each FCS thus has a value of $Q$. From these values we can compute the cumulative distribution of $Q$, that is, the probability that $Q$ is at least as large as a given value X. The resulting curve is shown here, along with the bisecting line. The intersection of the two shows that over 79% of the FCSs have a $Q$ value of at least 0.79. A randomization test of the hypothesis that reaction annotations are not correlated within the FCSs rejects this hypothesis at P<0.001 (see Methods for details).

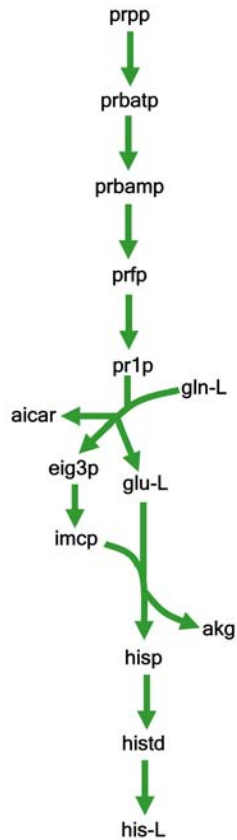

**Figure S3: Example of a frequently arising FCS in sampled genotypes; this FCS corresponds to a biochemical pathway.** We display here the most frequent FCS having more than 5 reactions. It occurs in 898 of the 1000 sampled genotypes with phenotype $V_{env}$ =89. All of its reactions belong to the histidine metabolism pathway. In this figure, the (hyper) edges represent reactions taking substrate metabolites to product metabolites (all these reactions are in fact irreversible). To reduce clutter, the ubiquitous high degree metabolites such as ATP, NADH, etc. have been omitted from this figure. Abbreviations of metabolite names are as follows: aicar = 5-Amino-1-(5-Phospho-D-ribosyl)imidazole-4-carboxamide; akg = 2-Oxoglutarate; eig3p = D-erythro-1-(Imidazol-4-yl)glycerol 3-phosphate; gln-L = L-Glutamine; glu-L = L-Glutamate; his-L = L-Histidine; hisp = L-Histidinol phosphate; histd = L-Histidinol; imacp = 3-(Imidazol-4-yl)-2-oxopropyl phosphate; prbamp = 1-(5-Phosphoribosyl)-AMP; prbatp = 1-(5-Phosphoribosyl)-ATP; prfp = 1-(5-Phosphoribosyl)-5-[(5-phosphoribosylamino)methylideneamino]imidazole-4-carboxamide; prlp = 5-[(5-phospho-1-deoxyribulos-1-ylamino)methylideneamino]-1-(5-phosphoribosyl)imidazole-4-carboxamide; prpp = 5-Phospho-alpha-D-ribose 1-diphosphate.

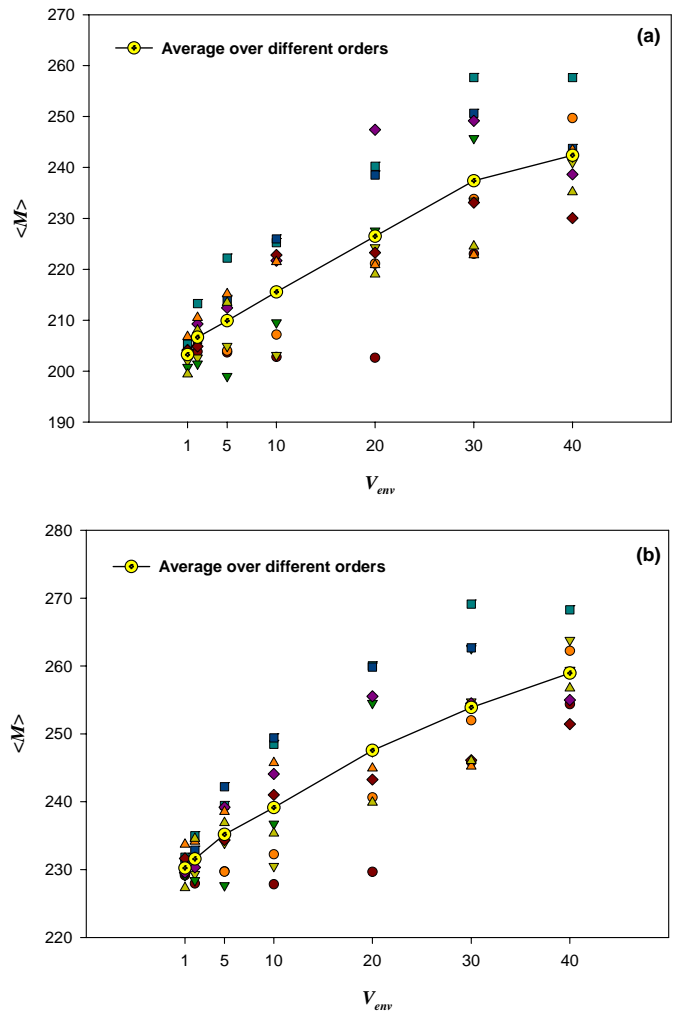

**Figure S4: Environmental versatility enhances the modularity index *M* regardless of the value of *n*.** Similar to what is shown in Fig. 2a, we display here the trend for the mean of the modularity index *M* as a function of increasing $V_{env}$, *i.e.*, as one forces genotypes to grow on more and more environments. **a)** Networks containing $n=500$ reactions. **b)** Networks containing $n=700$ reactions. We find the same qualitative increase in the average of *M* as $V_{env}$ increases, irrespective of the value of *n*.

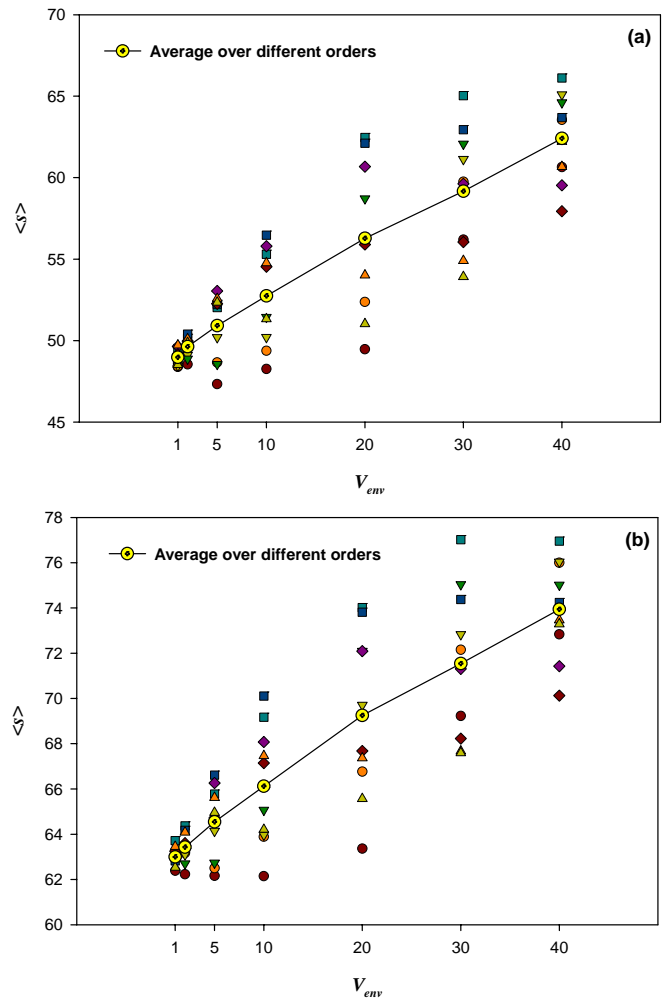

**Figure S5: Environmental versatility enhances the number of modules *s* regardless of the value of *n*.** Similar to what is shown in Fig. 2b, we display here the trend for the mean of *s* (the number of FCSs) as a function of increasing $V_{env}$, *i.e.*, as one forces genotypes to grow on more and more environments. **a)** $n=500$ reactions. **b)** $n=700$ reactions. We find the same increase in the average of *s* as $V_{env}$ increases, irrespective of the value of *n*.

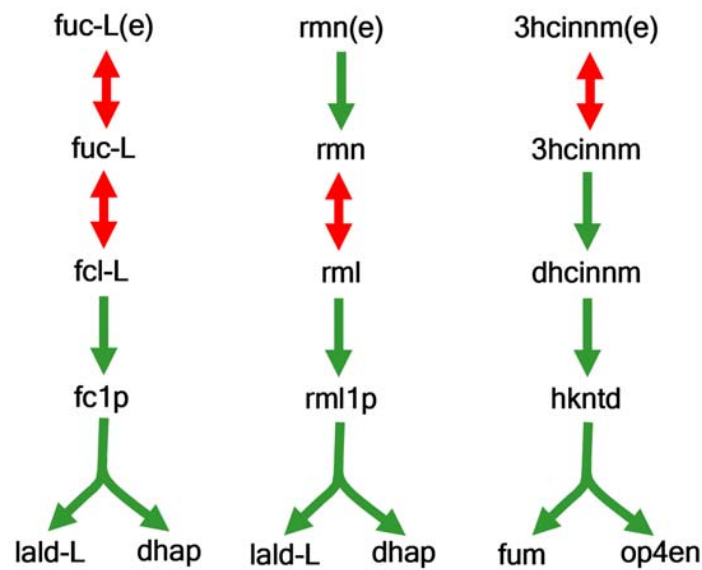

**Figure S6: Example of FCSs that account for the increase in modularity with versatility.** The set $R_{89}\backslash R_1$ is constructed from FCSs that are over-represented at high versatility. The set contains 24 reactions that are as close as possible to nutrients (have a scope distance of 1). Starting with all the FCSs in our 1000 sampled genotypes, we selected those that contained at least one of the 24 reactions. Shown here are the corresponding most frequent FCSs subject to the constraint of containing 4 or more reactions. In this figure, the (hyper) edges represent reactions involving metabolites. Green edges represent irreversible reactions and red edges represent reversible reactions. To reduce clutter, the ubiquitous high degree metabolites such as ATP, NADH, etc. have been omitted from this figure. Abbreviations of metabolite names are as follows: 3hcinnm = 3-hydroxycinnamic acid; 3hcinnm(e) = 3-hydroxycinnamic acid (Extracellular); dhap = Dihydroxyacetone phosphate; dhcinnm = 2,3-dihydroxicinnamic acid; fc1p = L-Fuculose 1-phosphate; fcl-L = L-fuculose; fuc-L = L-Fucose; fuc-L(e) = L-Fucose (Extracellular); fum = Fumarate; hkntd = 2-hydroxy-6-ketononatrienedioate; lald-L = L-Lactaldehyde; op4en = 2-Oxopent-4-enoate; rml = L-Rhamnulose; rml1p = L-Rhamnulose 1-phosphate; rmn = L-Rhamnose; rmn(e) = L-Rhamnose (Extracellular).

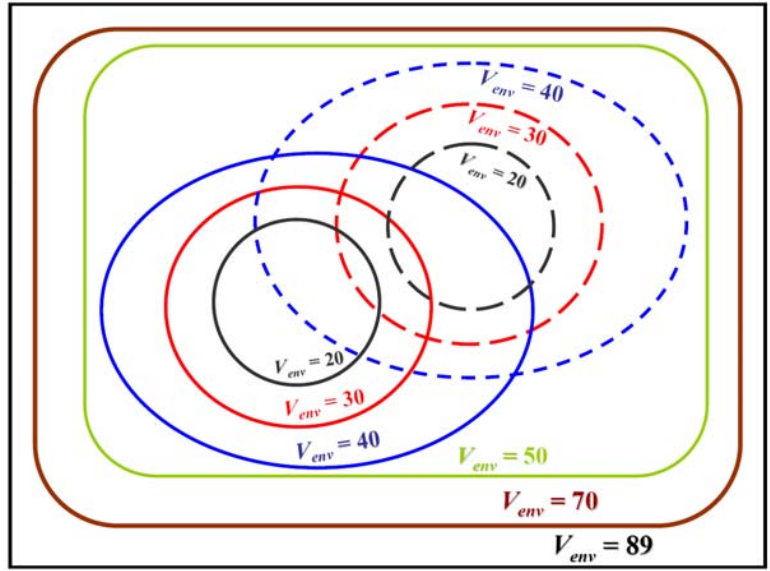

**Figure S7: Schematic diagram describing nested sets of environments used to sample genotypes with decreasing $V_{env}$.** Because the computations carried out here are very demanding in computing time, we have used only one subset of 70 environments within $V_{env}$ =89 to sample an ensemble with $V_{env}$ =70. Similarly, only one subset of 50 environments within the choice for $V_{env}$ =70 was used for sampling an ensemble with $V_{env}$ =50. However, for $V_{env}$ =40, we have taken 10 different subsets of 40 environments within the set used for $V_{env}$ =50; thus we generated samples in 10 different ensembles at $V_{env}$ =40. (Note that for clarity we have chosen to display only two different subsets for $V_{env}$ =40 in this figure). For $V_{env}$ =30, we used 10 subsets of 30 environments, one chosen from each of the sets used at $V_{env}$ =40, thereby sampling 10 different ensembles at $V_{env}$ =30. Similarly, for $V_{env}$ =20, we took 10 subsets of 20 environments, one from each of the 10 sets at $V_{env}$ =30, producing 10 different ensembles with $V_{env}$ =20. The same procedure was followed for $V_{env}$ =10, 5, 2 and 1. In summary, we took 10 different choices of nested environments to sample genotypes in the ensembles with $V_{env}$ = 1, 2, 5, 10, 20, 30 and 40, and averaged the properties of the sampled genotypes over the 10 different choices for these $V_{env}$ values. Note that the higher the phenotypic constraints are (the more environments we tested), the longer the computation times become for sampling genotypes in the ensemble of interest. That is why we have used just one set of environments for $V_{env}$ =50 and 70.

## Additional File 2: Tables S1 and S2

**Table S1: List of 89 minimal media.** Each carbon source is provided along with ammonia, iron, potassium, protons, pyrophosphate, sodium, sulfate, water and oxygen for uptake.

| Serial Number | Carbon Source |
|---|---|
| 1 | (S)-Propane-1,2-diol |
| 2 | 2-Dehydro-3-deoxy-D-gluconate |
| 3 | 2-Oxoglutarate |
| 4 | 3-(3-hydroxy-phenyl)propionate |
| 5 | 3-hydroxycinnamic acid |
| 6 | 4-Aminobutanoate |
| 7 | Acetaldehyde |
| 8 | Acetate |
| 9 | Acetoacetate |
| 10 | Adenosine |
| 11 | Allantoin |
| 12 | Butyrate (n-C4:0) |
| 13 | Citrate |
| 14 | Cytidine |
| 15 | D-Alanine |
| 16 | Deoxyadenosine |
| 17 | Deoxycytidine |
| 18 | Deoxyguanosine |
| 19 | Deoxyinosine |
| 20 | Deoxyuridine |
| 21 | D-Fructose |
| 22 | D-Galactarate |
| 23 | D-Galactonate |
| 24 | D-Galactose |
| 25 | D-Galacturonate |
| 26 | D-Glucarate |
| 27 | D-Gluconate |
| 28 | D-Glucosamine |
| 29 | D-Glucose |
| 30 | D-Glucose 6-phosphate |
| 31 | D-Glucuronate |
| 32 | D-Glyceraldehyde |
| 33 | Dihydroxyacetone |
| 34 | D-Lactate |
| 35 | D-Mannitol |
| 36 | D-Mannose |
| 37 | D-Mannose 6-phosphate |
| 38 | D-Ribose |
| 39 | D-Serine |
| 40 | D-Sorbitol |
| 41 | D-Xylose |
| 42 | Ethanol |
| 43 | Fumarate |
| 44 | Galactitol |
| 45 | Glycerol |
| 46 | Glycerol 3-phosphate |
| 47 | Glycine |
| 48 | Glycolate |
| 49 | Guanosine |
| 50 | Hexadecanoate (n-C16:0) |
| 51 | Inosine |
| 52 | Lactose |
| 53 | L-Alanine |
| 54 | L-Arabinose |
| 55 | L-Arginine |
| 56 | L-Asparagine |
| 57 | L-Asparate |
| 58 | L-Fucose |
| 59 | L-Glutamate |
| 60 | L-Glutamine |
| 61 | L-idonate |
| 62 | L-Lactate |
| 63 | L-Malate |
| 64 | L-Proline |
| 65 | L-Rhamnose |
| 66 | L-Serine |
| 67 | L-tartrate |
| 68 | L-Threonine |
| 69 | L-Tryptophan |
| 70 | Maltohexaose |
| 71 | Maltopentaose |
| 72 | Maltose |
| 73 | Maltotetraose |
| 74 | Maltotriose |
| 75 | Melibiose |
| 76 | N-acetyl-D-glucosamine |
| 77 | N-Acetyl-D-mannosamine |
| 78 | N-Acetylneuraminate |
| 79 | Octadecanoate (n-C18:0) |
| 80 | Ornithine |
| 81 | Phenylpropanoate |
| 82 | Putrescine |
| 83 | Pyruvate |
| 84 | Succinate |
| 85 | Sucrose |
| 86 | Tetradecanoate (n-C14:0) |
| 87 | Trehalose |
| 88 | Uridine |
| 89 | Xanthosine |

**Table S2: List of autocatalytic metabolites.** Following Kun et al (Genome Biology, 9:R51 (2008)), we added the listed autocatalytic metabolites (mostly cofactors) to the seed set of nutrient metabolites while computing distance using the scope algorithm.

| Serial Number | Metabolite name |
|---|---|
| 1 | 2-Demethyl menaquinone |
| 2 | Acyl carrier protein |
| 3 | ADP |
| 4 | Aldehyde |
| 5 | ATP |
| 6 | Biotin |
| 7 | CoA |
| 8 | CTP |
| 9 | Isopentenyl diphosphate |
| 10 | NAD+ |
| 11 | NADH |
| 12 | NADP+ |
| 13 | NADPH |
| 14 | Pyridoxal phosphate |
| 15 | Reduced ferredoxin |
| 16 | Theoredoxin |
| 17 | Thiamin |
| 18 | Ubiquinol-8 |
| 19 | Ubiquinone-8 |
| 20 | Undecaprenyl diphosphate |